 \documentclass[aps,preprint,showpacs]{revtex4}

\usepackage{amsmath}
\usepackage{amssymb}
\usepackage{graphicx}
\usepackage{epsfig}
\usepackage{subfigure}
\usepackage{rotating}
\usepackage{longtable}
\usepackage{multirow}
\usepackage{array} 
\usepackage{hyperref} 
\begin{document}
\title{ Study of Energy Spectra and Electromagnetic moments of Double Beta Decay Nuclei in Deformed Hartree-Fock Model}

\author{S. K. Ghorui$^{1,}$\footnote{email:surja@phy.iitkgp.ernet.in}, P. K. Raina$^{1,2}$, 
A. K. Singh$^{1}$, P. K. Rath$^{3}$, C. R. Praharaj$^{4}$}

\affiliation{$^{1}$ Department of Physics \& Meteorology, IIT Kharagpur, Kharagpur-721302, INDIA}
\affiliation{$^{2}$ Department of Physics, IIT Ropar, Rupnagar-140001, INDIA}
\affiliation{$^{3}$ Department of Physics, University of Lucknow, Lucknow-226007, INDIA}
\affiliation{$^{4}$ Institute of Physics, Bhubaneswar-751005, INDIA}


\begin{abstract}
Spectroscopy of nuclei in the mass range $A=76$ to $A=130$, participating in double 
beta decay processes are studied in the 
framework of the self-consistent deformed Hartree-Fock (HF) and angular momentum (J) 
projection model. Spectra of ground bands have been 
studied and compared with available experimental results for even-even parent and 
daughter as well as for intermediate odd-odd nuclei.  To test the reliability of the
wave functions we have also calculated the reduced E2 transition matrix elements, 
electric quadrupole moments and magnetic dipole moments for these nuclei. The 
calculated results are compared with the experimental findings and substantial 
agreement is achieved.  
\end{abstract}


\maketitle

\section{Introduction}
One of the most rarer processes of the nature, the double beta decay has unambiguous
importance in explicitly linking nuclear structure aspects with neutrino physics 
\cite{boh92,hax84,moe94}. Nuclear double beta decay is a second order process which 
involved electroweak decay of two nucleons simultaneously. The two neutrino mode of 
double beta decay is a second order process and allowed in standard model. It has been 
detected for nearly a dozen of nuclei \cite{ver02}. The correct theoretical description 
of these observations serves as a test of various nuclear models and also a necessary 
step to understand the neutrinoless mode. The extraction of information is essentially 
dependent upon the prediction of Nuclear Transition Matrix Elements (NTMEs) involved in 
these processes. The difference in the configurations of the nucleons between initial 
and final states is a major ingredient of the matrix elements. The NTMEs contain the wave 
function of the initial even-even nucleus in its $0^{+}$ ground state and the wave 
function of the final nucleus (usually $0^{+}$ ground state but some time also
excited $2^{+}$ and $0^{+}$ states). These wave functions are connected by proper operators.
In case of two neutrino mode the summation over all $1^{+}$ states of the intermediate
odd-odd nuclei is required where as for neutrinoless decay the summation over the complete
set of the states of intermediate nucleus is needed. Therefore it will be necessary to 
evaluate the structure of the initial, final and intermediate nuclei explicitly.

The most of the nuclei undergoing double beta decay are medium or heavy mass region and
all of them are even even type, in which pairing degrees of freedom play an important role.
Moreover it has been conjectured that the deformation plays a crucial role in double beta 
transitions. Hence it is desirable to have a model in which pairing and deformation degrees
of freedom treated on equal footing in its formalism. The shell model, which attempts to
solve the nuclear many-body problems as exactly as possible, is always being the best choice
for model calculations. However, for most of the double beta decaying nuclei the full scale 
shell model calculations are not feasible and the reliability of theoretical predictions has 
been hampered by unstabilities in case of BCS+RPA type treatment. For this purpose the angular 
momentum projected deformed Hartree-Fock (DHF) model is one of the most convenient choices. 
The DHF formalism, the pairing interaction is taken into account by mixing particle-hole 
excitations (2p-2h,4p-4h etc) of $K=0$ type across the Fermi surfaces over the HF ground state 
and deformation effects are included in a self-consistent way. Moreover it has been shown 
\cite{kha71} that it is possible to get shell model results with angular momentum projected 
DHF calculation by mixing only a few intrinsic configurations built by suitable particle-hole 
excitations across proton and neutron Fermi surfaces.

Here we have used microscopic many-body self consistent model based on deformed Hartree-Fock 
procedure and subsequent angular momentum projection technique for a reliable description of 
the nuclear structure of some medium as well heavy nuclei participating in double beta decay 
processes in the mass range $A=76$ to $130$. Since our aim is to study the low-lying 
spectroscopic properties of the nuclei, we have not incorporated the Coulomb effect
in our calculations. The inclusion of Coulomb effect will not change the energy spectra and 
electromagnetic matrix elements of individual nucleus rather it will change the energy 
difference between two nuclei.

In the present article we briefly discuss about the theoretical formalism in Section 2. 
Section 3 contains the discussion about the results obtained from this paper. The 
conclusions are finally given in Section 4.

\section{Deformed Hartree-Fock  and Angular Momentum (J) Projection Method}
In this section we present briefly the model used for the microscopic calculations.
More details can be found in Refs.~\cite{pra82,rip66,pra99,skg11}. 
Our model consists of self-consistent deformed Hartree-Fock mean field obtained
with a Surface Delta residual interaction and subsequent Angular momentum projection
to obtain states with good angular momentum. 

The axially deformed states $|\eta m\rangle$ are expanded in the spherical basis states as follows:
\begin{equation}
|\eta m\rangle=\sum_{j}C_{jm}|jm\rangle
\end{equation}
where j is the angular momentum of the spherical single particle state and m its
projection on symmetry axis. The mixing amplitude $C_{jm}$ are obtained by solving 
deformed Hartree-Fock equations in an iterative process. When the convergence in the 
HF solutions is obtained we get deformed single particle orbits. The residual interaction
is also included self-consistently and it causes the mixing of  single-particle orbits of nucleons. 

 Because of mixing in single particle orbits, the HF configurations $|\phi_K\rangle$ are
superposition of states of good angular momentum. The states of good angular momentum can
be extracted by means of projection operator \cite{pei57}  

\begin{equation}
P_{K}^{JM} = \frac{2J+1}{8\pi^2}\int d\Omega \ {D_{MK}^J(\Omega)}^*
\ R(\Omega)
\label{eq_1}
\end{equation}
Here $R(\Omega)$ is the rotation operator 
$e^{-i\alpha J_z}e^{-i\beta J_y}e^{-i\gamma J_z}$
and $\Omega$ represents the Euler angles $\alpha$, $\beta$ and $\gamma$.

The Hamiltonian overlap between two states of angular 
momentum J projected from intrinsic states $\phi_{K_1}$ and  $\phi_{K_2}$ is given by:
\begin{eqnarray}
H_{K_1K_2}^J = \frac{2J+1}{2} \frac{1}{(N_{K_1K_1}^JN_{K_2K_2}^J)^{1/2}}
{{{\int}_{0}}^{\pi}} d\beta \ sin(\beta)d_{K_1K_2}^J(\beta)\\ \nonumber
\times
\langle\phi_{K_1}|He^{-i\beta J_y}|\phi_{K_2}\rangle
\label{eq_2}
\end{eqnarray}
with 
\begin{equation}
N_{K_1K_2}^J = \frac{2J+1}{2}{{{\int}_{0}}^{\pi}} d\beta \ sin(\beta)
d_{K_1K_2}^J(\beta)\langle\phi_{K_1}|e^{-i\beta J_y}|\phi_{K_2}\rangle
\label{eq_3}
\end{equation}

Reduced matrix elements of tensor operator T$^L$ between projected
states $\psi_{K1}^{J_1}$ and  $\psi_{K2}^{J_2}$ are given by
\begin{eqnarray}
\langle\psi_{K_1}^{J_1}||T^L||\psi_{K_2}^{J_2}\rangle = \frac{1}{2}
\frac{(2J_2+1)(2J_1+1)^{1/2}}{(N_{K_1K_1}^{J_1}N_{K_2K_2}^{J_2})^{1/2}}
\sum_{\mu\nu}C_{\mu\nu K_1}^{J_2LJ_1} \\ \nonumber
\times 
{{{\int}_{0}}^{\pi}} d\beta \ sin(\beta)d_{\mu K_2}^{J_2}(\beta)
\langle\phi_{K_1}|T^L_{\nu}e^{-i\beta J_y}|\phi_{K_2}\rangle
\label{eq_4}
\end{eqnarray}
where the tensor operator $T^L$ denotes electromagnetic operators of multipolarity L.

In general, two states $|{\psi_{K_1}^{JM}}\rangle$ and
$|{\psi_{K_2}^{JM}}\rangle$
projected from two intrinsic configurations $|\phi_{K_1}\rangle$ and
$|\phi_{K_2}\rangle$ are not orthogonal to each other even if
 the intrinsic states $|\phi_{K_1}\rangle$
and $|\phi_{K_2}\rangle$ are orthogonal. We orthonormalise them
 using following  equation
\begin{equation}
\sum_{K^{\prime}}(H^J_{KK^{\prime}}-E_JN^J_{KK^{\prime}})
b^J_{K^{\prime}}=0
\end{equation}
Here $N^J_{KK^{\prime}}$ are amplitude overlap and $b^J_{K^{\prime}}$ are
the orthonormalised amplitudes, which can be identified as
band-mixing amplitudes. The orthonormalised states are given by
\begin{equation}
\psi^{JM}=\sum_{K}b^J_{K}\psi_{K}^{JM}
\end{equation}
With these orthonormalised states we can calculate matrix elements of
various tensor operators.

	The spectroscopic quadrupole moment of a state with angular momentum J is given by

\begin{equation}
Q_{S}(J)=\frac{1}{(2J+1)^{1/2}}\left(\frac{16\pi}{5}\right)^{1/2}C^{J2J}_{J0J}
\langle \Psi^{J}_{K}||\sum_{i=p,n}Q_{2}^{i} ||\Psi^{J}_{K}\rangle 
\end{equation}
where the summation is for quadrupole moment operators of protons and neutrons. 

The gyromagnetic factor (g-factor) of state J is defined as
\begin{equation}
g(J)=\frac{\mu(J)}{J}
\end{equation}
where $\mu(J)$ is the magnetic moment of state J.
The magnetic dipole moment ($\mu$) of a state with angular momentum J can be expressed as
\begin{equation}
\mu(J)=\frac{1}{(2J+1)^{1/2}}C^{J1J}_{J0J}
\left(\sum_{i=p,n}\langle \Psi^{J}_{K}|| g_{l}^{i}{l}_{i} + g_{s}^{i}{s}_{i}||
\Psi^{J}_{K}\rangle \right)
\end{equation}
where i=p and n for protons and neutrons respectively. 
where $g_{l}$ and $g_{s}$ are orbital and spin g-factor respectively. The g-factors of 
$g_{l}=1.0 \mu_{N}$ and $g_{s}=5.586\times0.5 \mu_{N} $ for protons and 
$g_{l}=0 \mu_{N}$ and $g_{s}=-3.826\times0.5 \mu_{N}$ for neutrons are used for the
calculations. The quenching of spin g-factors by 0.75 is taken in account to consider
the core polarization effect \cite{nil61,boch65}.

\section{Results and Discussion}
 The deformed HF orbits are calculated with a spherical core of $^{56}$Ni, the model 
space spans the 1$p_{3/2}$, 0$f_{5/2}$, 1$p_{1/2}$, 0$g_{9/2}$, 0$d_{5/2}$, 
0$g_{7/2}$, 0$d_{3/2}$, 2$s_{1/2}$ and 0$h_{11/2}$ orbits both for protons neutrons with 
single particle energies 0.0, 0.78, 1.88, 4.44, 8.88, 11.47, 10.73, 12.21 and 13.69 MeV 
respectively. We use a surface delta interaction \cite{fae67} ( with interaction strength 
$\sim$0.36 for $p-p$, $p-n$ and $n-n$ interactions) as the residual interaction among the 
active nucleons in these orbits. 

Deformed Hartree-Fock and Angular Momentum Projection calculations are performed for some
medium-heavy nuclei with mass number $A=76$ to $A=130$. In our model we can calculate 
the energy spectra and other electromagnetic moments for even-even parents and daughter 
as well as odd-odd intermediate nuclei. This is an interesting feature of our model.

In contrast to PSM where a large number of configurations is needed to understand low-lying
yrast spectra, it is found that angular momentum projection (AMP) from a few low-lying 
configurations can reasonably reproduces the yrast-spectra. Sometimes the AMP from single K 
configuration gives fairly good description of yrast structure. This is due to fact that 
the residual interaction is included self-consistently in the HF calculations, so that
the HF configurations and various particle-hole configurations built on HF solutions are
already closer to the solutions obtained from full many body Schr\"{o}dinger equations. 

The Hartree-Fock solutions for the parent, intermediate and daughter nuclei are given
in Table~\ref{tab:tab_hf-sol}. The values of intrinsic quadrupole moments(Q$_{20}$)
are given both for proton and neutron in unit of harmonic oscillator length parameter,
b ($=0.9A^{1/3}+0.7$fm).
\begin{longtable}{llllll}
\caption{Hatree-Fock Solution of the Nuclei}\\
\hline\hline
Nucleus & Shape & E$_{HF}$ &  & \multicolumn{2}{c}{$\left\langle
Q_{20}\right\rangle $ (in b$^{2}$)} \\ \cline{5-6}
&  & \multicolumn{1}{l}{ (in MeV)} & \multicolumn{1}{l}{} & Proton & 
Neutron \\ 
\hline
\endfirsthead
\multicolumn{6}{c}{{\tablename} \thetable{} -- Continued} \\
\hline
Nucleus & Shape & E$_{HF}$ &  & \multicolumn{2}{c}{$\left\langle
Q_{20}\right\rangle $ (in b$^{2}$)} \\ \cline{5-6}
&  & \multicolumn{1}{l}{ (in MeV)} & \multicolumn{1}{l}{} & Proton & 
Neutron \\ 

\hline
\endhead
\hline
\multicolumn{6}{l}{{Continued on next page\ldots}} \\
\endfoot
\endlastfoot

$^{76}$Ge & Prolate & \multicolumn{1}{l}{-45.772} & \multicolumn{1}{l}{} & 
\multicolumn{1}{r}{3.964} & \multicolumn{1}{r}{4.659} \\ 
& Oblate & \multicolumn{1}{l}{-43.672} & \multicolumn{1}{l}{} & 
\multicolumn{1}{r}{-2.995} & \multicolumn{1}{r}{-3.439} \\ 
$^{76}$As & Prolate & \multicolumn{1}{l}{-48.909} & \multicolumn{1}{l}{} & 
\multicolumn{1}{r}{4.181} & \multicolumn{1}{r}{6.712} \\ 
& Oblate & \multicolumn{1}{l}{-46.819} & \multicolumn{1}{l}{} & 
\multicolumn{1}{r}{-3.675} & \multicolumn{1}{r}{-2.952} \\ 
$^{76}$Se & Prolate & \multicolumn{1}{l}{-55.613} & \multicolumn{1}{l}{} & 
\multicolumn{1}{r}{4.699} & \multicolumn{1}{r}{8.894} \\ 
& Oblate & \multicolumn{1}{l}{-54.573} & \multicolumn{1}{l}{} & 
\multicolumn{1}{r}{-4.582} & \multicolumn{1}{r}{-2.518} \\ 
$^{78}$Kr & Prolate & \multicolumn{1}{l}{-72.058} & \multicolumn{1}{l}{} & 
\multicolumn{1}{r}{8.573} & \multicolumn{1}{r}{10.372} \\ 
& Oblate & \multicolumn{1}{l}{-71.054} & \multicolumn{1}{l}{} & 
\multicolumn{1}{r}{-7.274} & \multicolumn{1}{r}{-6.739} \\ 
$^{78}$Br & Prolate & \multicolumn{1}{l}{-68.389} & \multicolumn{1}{l}{} & 
\multicolumn{1}{r}{8.259} & \multicolumn{1}{r}{9.599} \\ 
& Oblate & \multicolumn{1}{l}{-67.104} & \multicolumn{1}{l}{} & 
\multicolumn{1}{r}{-4.634} & \multicolumn{1}{r}{-2.978} \\ 
$^{78}$Se & Prolate & \multicolumn{1}{l}{-59.584} & \multicolumn{1}{l}{} & 
\multicolumn{1}{r}{4.711} & \multicolumn{1}{r}{7.640} \\ 
& Oblate & \multicolumn{1}{l}{-57.790} & \multicolumn{1}{l}{} & 
\multicolumn{1}{r}{-4.565} & \multicolumn{1}{r}{-3.452} \\ 
$^{82}$Se & Prolate & \multicolumn{1}{l}{-64.588} & \multicolumn{1}{l}{} & 
\multicolumn{1}{r}{4.648} & \multicolumn{1}{r}{4.865} \\ 
& Oblate & \multicolumn{1}{l}{-61.932} & \multicolumn{1}{l}{} & 
\multicolumn{1}{r}{-4.626} & \multicolumn{1}{r}{--4.020} \\ 
$^{82}$Br & Prolate & \multicolumn{1}{l}{-68.428} & \multicolumn{1}{l}{} & 
\multicolumn{1}{r}{3.821} & \multicolumn{1}{r}{5.129} \\ 
& Oblate & \multicolumn{1}{l}{-66.674} & \multicolumn{1}{l}{} & 
\multicolumn{1}{r}{-4.297} & \multicolumn{1}{r}{-3.794} \\ 
$^{82}$Kr & Prolate & \multicolumn{1}{l}{-77.646} & \multicolumn{1}{l}{} & 
\multicolumn{1}{r}{3.210} & \multicolumn{1}{r}{4.789} \\ 
& Oblate & \multicolumn{1}{l}{-77.107} & \multicolumn{1}{l}{} & 
\multicolumn{1}{r}{-3.767} & \multicolumn{1}{r}{-3.330} \\ 
$^{96}$Mo & Prolate & \multicolumn{1}{l}{-91.214} & \multicolumn{1}{l}{} & 
\multicolumn{1}{r}{3.745} & \multicolumn{1}{r}{5.820} \\ 
& Oblate & \multicolumn{1}{l}{-90.418} & \multicolumn{1}{l}{} & 
\multicolumn{1}{r}{-2.518} & \multicolumn{1}{r}{-3.445} \\ 
$^{96}$Tc & Prolate & \multicolumn{1}{l}{-98.572} & \multicolumn{1}{l}{} & 
\multicolumn{1}{r}{4.562} & \multicolumn{1}{r}{5.456} \\ 
& Oblate & \multicolumn{1}{l}{-96.403} & \multicolumn{1}{l}{} & 
\multicolumn{1}{r}{-2.976} & \multicolumn{1}{r}{-3.361} \\ 
$^{96}$Ru & Prolate & \multicolumn{1}{l}{-106.457} & \multicolumn{1}{l}{} & 
\multicolumn{1}{r}{5.951} & \multicolumn{1}{r}{6.037} \\ 
& Oblate & \multicolumn{1}{l}{-103.387} & \multicolumn{1}{l}{} & 
\multicolumn{1}{r}{-3.406} & \multicolumn{1}{r}{-3.795} \\ 
$^{100}$Mo & Prolate & \multicolumn{1}{l}{-107.174} & \multicolumn{1}{l}{} & 
\multicolumn{1}{r}{3.694} & \multicolumn{1}{r}{5.390} \\ 
& Oblate & \multicolumn{1}{l}{-105.160} & \multicolumn{1}{l}{} & 
\multicolumn{1}{r}{-2.515} & \multicolumn{1}{r}{-1.677} \\ 
$^{100}$Tc & Prolate & \multicolumn{1}{l}{-118.232} & \multicolumn{1}{l}{} & 
\multicolumn{1}{r}{9.857} & \multicolumn{1}{r}{10.681} \\ 
& Oblate & \multicolumn{1}{l}{-112.926} & \multicolumn{1}{l}{} & 
\multicolumn{1}{r}{-3.004} & \multicolumn{1}{r}{-3.016} \\ 
$^{100}$Ru & Prolate & \multicolumn{1}{l}{-127.008} & \multicolumn{1}{l}{} & 
\multicolumn{1}{r}{6.338} & \multicolumn{1}{r}{8.198} \\ 
& Oblate & \multicolumn{1}{l}{-123.029} & \multicolumn{1}{l}{} & 
\multicolumn{1}{r}{-3.386} & \multicolumn{1}{r}{-3.399} \\ 
$^{106}$Cd & Prolate & \multicolumn{1}{l}{-172.047} & \multicolumn{1}{l}{} & 
\multicolumn{1}{r}{6.079} & \multicolumn{1}{r}{10.236} \\ 
& Oblate & \multicolumn{1}{l}{-167.235} & \multicolumn{1}{l}{} & 
\multicolumn{1}{r}{-9.462} & \multicolumn{1}{r}{-11.471} \\ 
$^{106}$Ag & Prolate & \multicolumn{1}{l}{-166.746} & \multicolumn{1}{l}{} & 
\multicolumn{1}{r}{6.581} & \multicolumn{1}{r}{9.761} \\ 
& Oblate & \multicolumn{1}{l}{-164.073} & \multicolumn{1}{l}{} & 
\multicolumn{1}{r}{-8.909} & \multicolumn{1}{r}{-10.966} \\ 
$^{106}$Pd & Prolate & \multicolumn{1}{l}{-159.401} & \multicolumn{1}{l}{} & 
\multicolumn{1}{r}{6.916} & \multicolumn{1}{r}{8.847} \\ 
& Oblate & \multicolumn{1}{l}{-156.593} & \multicolumn{1}{l}{} & 
\multicolumn{1}{r}{-6.728} & \multicolumn{1}{r}{-9.381} \\ 
$^{110}$Pd & Prolate & \multicolumn{1}{l}{-170.686} & \multicolumn{1}{l}{} & 
\multicolumn{1}{r}{6.079} & \multicolumn{1}{r}{6.754} \\ 
& Oblate & \multicolumn{1}{l}{-170.009} & \multicolumn{1}{l}{} & 
\multicolumn{1}{r}{-4.862} & \multicolumn{1}{r}{-7.360} \\ 
$^{110}$Ag & Prolate & \multicolumn{1}{l}{-183.424} & \multicolumn{1}{l}{} & 
\multicolumn{1}{r}{5.474} & \multicolumn{1}{r}{7.796} \\ 
& Oblate & \multicolumn{1}{l}{-181.197} & \multicolumn{1}{l}{} & 
\multicolumn{1}{r}{-11.114} & \multicolumn{1}{r}{-15.346} \\ 
$^{110}$Cd & Prolate & \multicolumn{1}{l}{-196.677} & \multicolumn{1}{l}{} & 
\multicolumn{1}{r}{5.621} & \multicolumn{1}{r}{9.471} \\ 
& Oblate & \multicolumn{1}{l}{-192.382} & \multicolumn{1}{l}{} & 
\multicolumn{1}{r}{-12.079} & \multicolumn{1}{r}{-14.254} \\ 
$^{116}$Cd & Prolate & \multicolumn{1}{l}{-202.919} &  & \multicolumn{1}{r}{
6.337} & \multicolumn{1}{r}{10.633} \\ 
& Oblate & \multicolumn{1}{l}{-200.953} &  & \multicolumn{1}{r}{-2.575} & 
\multicolumn{1}{r}{-4.562} \\ 
$^{116}$In & Prolate & \multicolumn{1}{l}{-211.305} &  & \multicolumn{1}{r}{
10.883} & \multicolumn{1}{r}{13.402} \\ 
& Oblate & \multicolumn{1}{l}{-210.156} &  & \multicolumn{1}{r}{-11.879} & 
\multicolumn{1}{r}{-15.857} \\ 
$^{116}$Sn & Prolate & \multicolumn{1}{l}{-220.357} &  & \multicolumn{1}{r}{
1.969} & \multicolumn{1}{r}{4.177} \\ 
& Oblate & \multicolumn{1}{l}{-219.197} &  & \multicolumn{1}{r}{-1.183} & 
\multicolumn{1}{r}{-5.466} \\ 
$^{124}$Sn & Prolate & \multicolumn{1}{l}{-229.001} & \multicolumn{1}{l}{} & 
\multicolumn{1}{r}{2.899} & \multicolumn{1}{r}{3.670} \\ 
& Oblate & \multicolumn{1}{l}{-228.494} & \multicolumn{1}{l}{} & 
\multicolumn{1}{r}{-2.147} & \multicolumn{1}{r}{-4.587} \\ 
$^{124}$Sb & Prolate & \multicolumn{1}{l}{-237.729} & \multicolumn{1}{l}{} & 
\multicolumn{1}{r}{2.575} & \multicolumn{1}{r}{2.839} \\ 
& Oblate & \multicolumn{1}{l}{-240.753} & \multicolumn{1}{l}{} & 
\multicolumn{1}{r}{-5.044} & \multicolumn{1}{r}{-9.165} \\ 
$^{124}$Te & Prolate & \multicolumn{1}{l}{-246.776} & \multicolumn{1}{l}{} & 
\multicolumn{1}{r}{3.652} & \multicolumn{1}{r}{2.008} \\ 
& Oblate & \multicolumn{1}{l}{-250.869} & \multicolumn{1}{l}{} & 
\multicolumn{1}{r}{-5.649} & \multicolumn{1}{r}{-9.108} \\ 
$^{130}$Te & Prolate & \multicolumn{1}{l}{-267.880} & \multicolumn{1}{l}{} & 
\multicolumn{1}{r}{4.737} & \multicolumn{1}{r}{4.587} \\ 
& Oblate & \multicolumn{1}{l}{-268.664} & \multicolumn{1}{l}{} & 
\multicolumn{1}{r}{-3.716} & \multicolumn{1}{r}{-3.670} \\ 
$^{130}$I & Prolate & \multicolumn{1}{l}{-273.928} & \multicolumn{1}{l}{} & 
\multicolumn{1}{r}{4.806} & \multicolumn{1}{r}{4.616} \\ 
& Oblate & \multicolumn{1}{l}{-274.691} & \multicolumn{1}{l}{} & 
\multicolumn{1}{r}{-3.896} & \multicolumn{1}{r}{-7.477} \\ 
$^{130}$Xe & Prolate & \multicolumn{1}{l}{-291.346} & \multicolumn{1}{l}{} & 
\multicolumn{1}{r}{5.105} & \multicolumn{1}{r}{7.570} \\ 
& Oblate & \multicolumn{1}{l}{-291.945} & \multicolumn{1}{l}{} & 
\multicolumn{1}{r}{-3.891} & \multicolumn{1}{r}{-8.331} \\ \hline\hline
\label{tab:tab_hf-sol}
\end{longtable}
\subsection{Energy Spectra}

\begin{figure}[hbt]
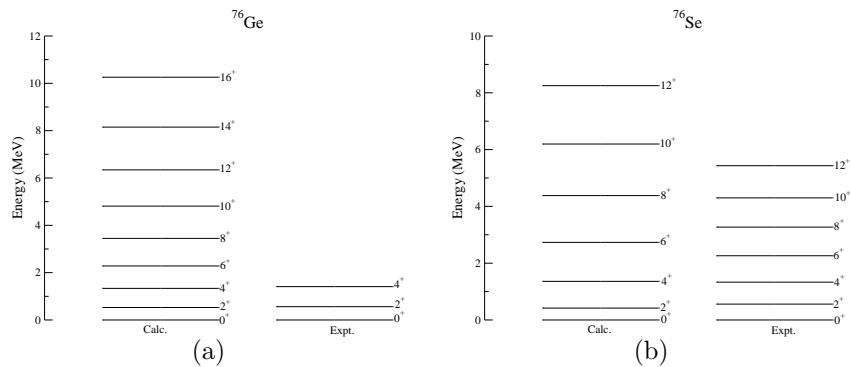

\begin{center}
\mbox{
\subfigure[]{\includegraphics[scale=0.25, angle=-90]{plots/spectra-Ge76.epsi}}
\hspace{ 0.5cm}
\subfigure[]{\includegraphics[scale=0.25, angle=-90]{plots/spectra-Se76.epsi}}}
\caption{Energy spectra for $^{76}$Ge and $^{76}$Se nuclei. The experimental values are 
taken from Ref. \cite{nds76}}
\label{fig:spec_ge-se}
\end{center}
\end{figure}

	In Fig.~\ref{fig:spec_ge-se} we have shown the low-lying spectra for the nuclei 
participating in the double beta decay processes of $^{76}$Ge. In case of the odd-odd 
intermediate nucleus $^{76}$As, the experimental energy spectra is not known therefore we are also not 
showing the theoretical spectra for these nuclei. For the nuclei $^{76}$Ge and $^{76}$Se 
we have considered the prolate solutions as the prolate solutions are energetically lower 
than oblate solutions. The angular momentum projected results in comparison with the 
available experimental data are shown in Figure~\ref{fig:spec_ge-se}. \\
\begin{figure}[hbt]
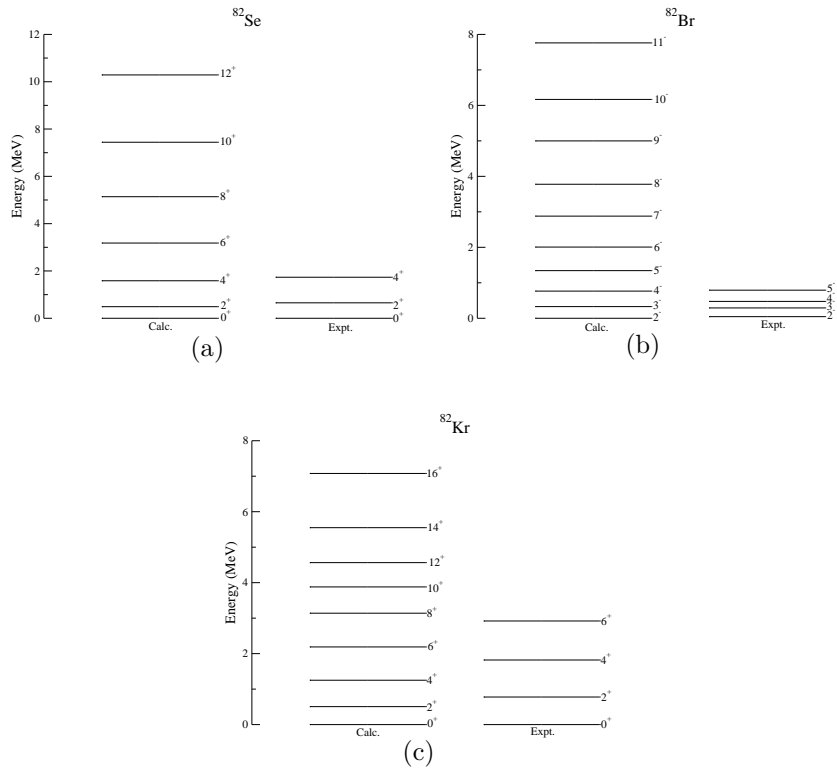

\begin{center}
\mbox{
\subfigure[]{\includegraphics[scale=0.25, angle=-90]{plots/spectra-Se82.epsi}}
\hspace{ 0.5cm}
\subfigure[]{\includegraphics[scale=0.25, angle=-90]{plots/spectra-Br82.epsi}}}\\
\vspace{ 0.3cm}
\subfigure[]{\includegraphics[scale=0.25, angle=-90]{plots/spectra-Kr82.epsi}}
\caption{Energy spectra for $^{82}$Se,$^{82}$Br and $^{82}$Kr nuclei. The experimental values are 
taken from Ref. \cite{nds82}}
\label{fig:spec_se-kr}
\end{center}
\end{figure}

The energy spectra of $^{82}$Se, $^{82}$Br and $^{82}$Kr are shown in Fig.~\ref{fig:spec_se-kr}.
The angular momentum projection results considering the prolate solutions of $^{82}$Se and $^{82}$Br
are shown in Fig.~\ref{fig:spec_se-kr}(a) and Fig.~\ref{fig:spec_se-kr}(b).
In case of $^{82}$Kr, the prolate and oblate solutions are degenerate. So for this nucleus we 
have performed the shape-mixing calculation by mixing the prolate and oblate configurations. 
The shape-mixing calculation reasonably reproduces the low lying spectra for $^{82}$Kr as can be 
seen in Fig.~\ref{fig:spec_se-kr}(c). \\

\begin{figure}[htp]
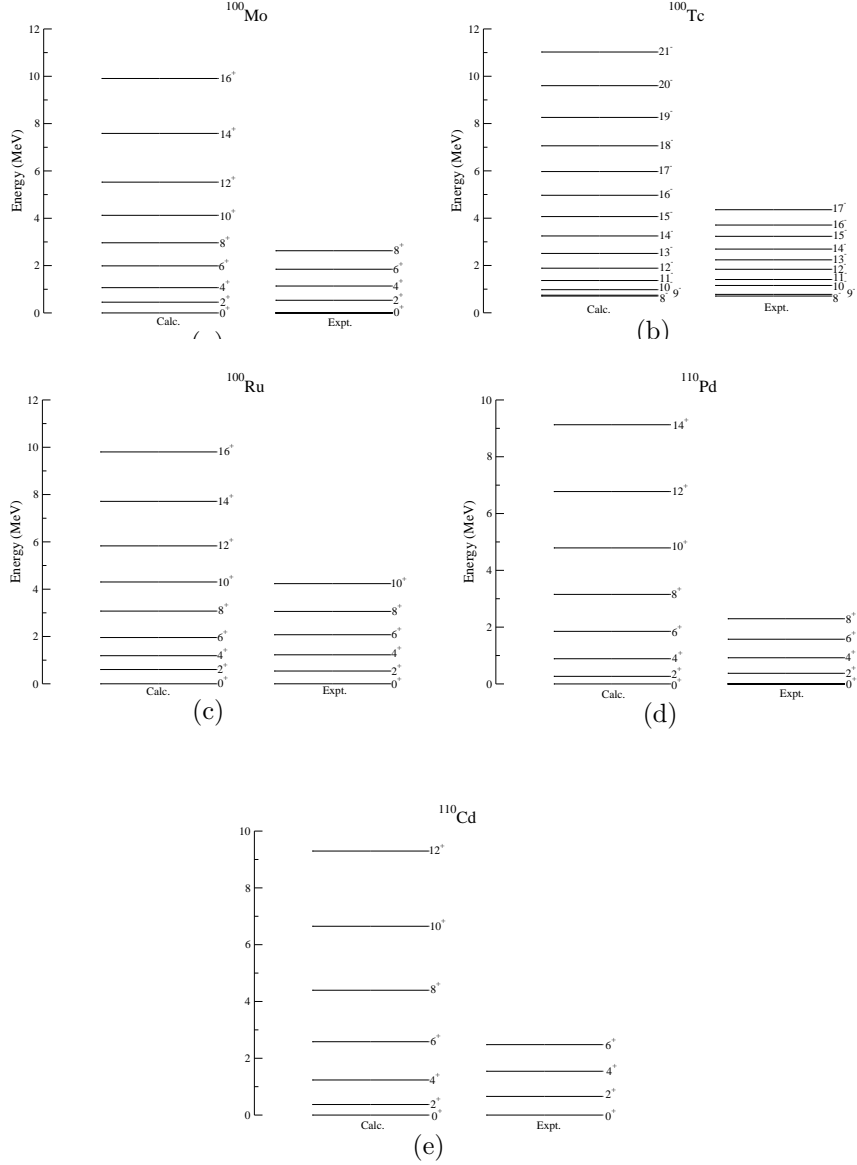

\centering
\mbox{
\subfigure[]{\includegraphics[scale=0.25, angle=-90]{plots/spectra-Mo100.epsi}}
\hspace{ 0.5cm}
\subfigure[]{\includegraphics[scale=0.25, angle=-90]{plots/spectra-Tc100.epsi}}}
\vspace{ 0.8cm}
\mbox{
\subfigure[]{\includegraphics[scale=0.25, angle=-90]{plots/spectra-Ru100.epsi}}

\hspace{0.5cm}
\subfigure[]{\includegraphics[scale=0.25, angle=-90]{plots/spectra-Pd110.epsi}}}
\subfigure[]{\includegraphics[scale=0.25, angle=-90]{plots/spectra-Cd110.epsi}}
\caption{Energy spectra for $^{100}$Mo, $^{100}$Tc, $^{100}$Ru  $^{110}$Pd and $^{110}$Cd nuclei.
The experimental values are taken from Refs. \cite{nds100,nds110}}
\label{fig:spec_mo-pd-cd}
\end{figure}

The comparison of calculated and experimental results for nuclei $^{100}$Mo, $^{100}$Tc, $^{100}$Ru, $^{110}$Pd
and $^{110}$Cd are shown in Fig.~\ref{fig:spec_mo-pd-cd}. 
A fairly good agreement between theoretical
and experimental spectra is obtained for them.\\

\begin{figure}[htp]
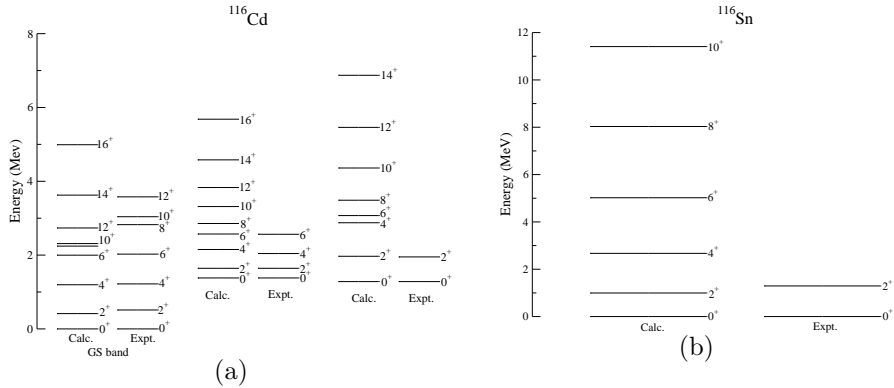

\centering
\mbox{
\subfigure[]{\includegraphics[scale=0.26, angle=-90]{plots/spectra-Cd116.epsi}}
\hspace{ 0.5cm}
\subfigure[]{\includegraphics[scale=0.25, angle=-90]{plots/spectra-Sn116.epsi}}}
\caption{Energy spectra for $^{116}$Cd and $^{116}$Sn nuclei. 
The experimental values are taken from Refs. \cite{nds116}}
\label{fig:spec_cd-sn}
\end{figure}

The nucleus $^{116}$Cd exhibits a complex structure as it has active protons near $ Z=50$
shell closure and active neutrons near the neutron midshell. Our self-consistent calculations
reproduce the band structure quit well. The compression in the ground band near $J=8\hbar$
is occurring due to the crossing of 2-proton excitation bands across $Z=50$ shell. 
Apart from the ground band we have calculated two excited $K=0^{+}$ bands compared with
experimental results in Fig.~\ref{fig:spec_cd-sn}(a). Similarly we have calculated the 
ground band for the daughter nucleus $^{116}$Sn and shown in Fig.~\ref{fig:spec_cd-sn}(b).\\

\begin{figure}[htp]
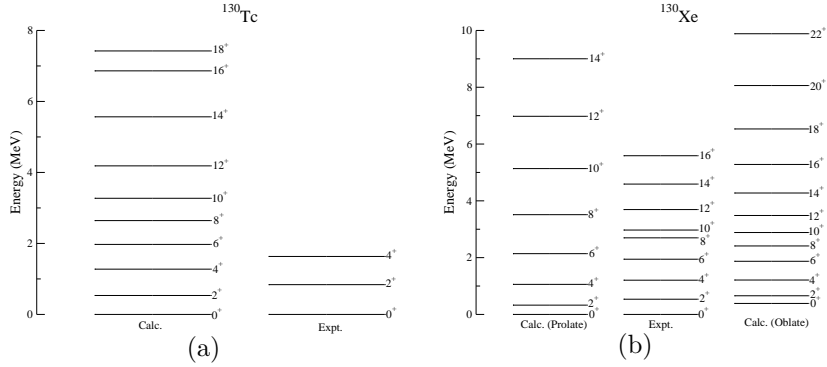

\centering
\mbox{
\subfigure[]{\includegraphics[scale=0.25, angle=-90]{plots/spectra-Te130.epsi}}
\hspace{ 0.5cm}
\subfigure[]{\includegraphics[scale=0.25, angle=-90]{plots/spectra-Xe130.epsi}}}
\caption{Energy spectra for $^{130}$Te and $^{130}$Xe nuclei. 
The experimental values are taken from Refs. \cite{nds130}}
\label{fig:spec_te-xe}
\end{figure}

  In Fig.~\ref{fig:spec_sn-te}(a) we have plotted the ground state spectra for
$^{124}$Sn. The Te isotopes are 2 protons above the tin shell closure. The ground
band spectra for  $^{124}$Te are shown in Figs.~\ref{fig:spec_sn-te}(b). In 
Fig.~\ref{fig:spec_te-xe}(a), the energy spectra of $^{130}$Te are given. 
We have used the oblate solutions for $^{128,130}$Te as they are lower in energy in
these two nuclei. The $^{130}$Xe isotope shows interesting feature. The interplay between
prolate and oblate shapes are observed in our calculations. The band spectra for 
$^{130}$Xe are shown in Fig~\ref{fig:spec_te-xe}(b). The low spin states of this 
isotope are of prolate shape but for the states above $J=6\hbar$ the calculations with
oblate shapes are more closer to the experimental levels.  \\ 

\begin{figure}[htp]
\centering
\mbox{
\subfigure[]{\includegraphics[scale=0.25, angle=-90]{plots/spectra-Sn124.epsi}}
\hspace{ 0.5cm}
\subfigure[]{\includegraphics[scale=0.25, angle=-90]{plots/spectra-Te124.epsi}}}
\caption{Energy spectra for $^{124}$Sn and $^{124}$Te nuclei. The experimental
 values are taken from Refs. \cite{nds124}}
\label{fig:spec_sn-te}
\end{figure}

The nuclei $^{78}$Kr, $^{96}$Ru and $^{106}$Cd are the most promising nuclei among
six which can decay through all three possible channels of positron double beta decay.
In our present study we have considered these double beta decay nuclei to study the
low-lying spectroscopy properties.\\

\begin{figure}[hbt]
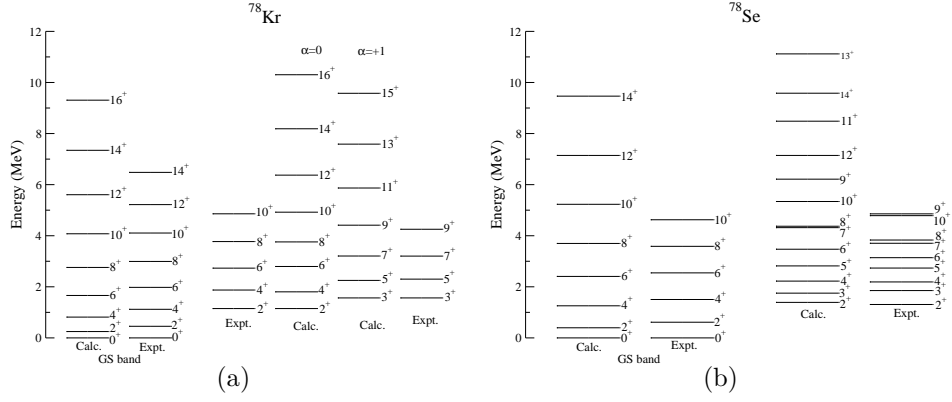

\centering
\subfigure[]{\includegraphics[scale=0.27, angle=-90]{plots/spectra-Kr78.epsi}}
\hspace{ 0.2cm}
\subfigure[]{\includegraphics[scale=0.27, angle=-90]{plots/spectra-Se78.epsi}}
\caption{Energy spectra for $^{78}$Kr and $^{78}$Se  nuclei
The experimental values are taken from Refs. \cite{nds78}}
\label{fig:spec_kr78}
\end{figure}

For $^{78}$Kr, the energy spectra are shown in Fig.~\ref{fig:spec_kr78}(a). 
Apart from the ground band we have shown the excited $K=2^{+}$ band. The signature
effect is observed in this band due to the rotational alignment of $\nu$g$_{9/2}$.
Similarly for $^{78}$Se nucleus, the ground and excited $K=2^{+}$ bands are shown
in Fig.~\ref{fig:spec_kr78}(b). For the $K=2^{+}$ band the signature inversion
is observed as a result of band crossing between $\nu$f$_{5/2}$$\nu$g$_{9/2}$
decoupled band with $K=2^{+}$ and $K=1^{+}$. In these nuclei the rotational
behaviour of the ground band reproduces quite well except the $2^{+}\rightarrow 0^{+}$
separation. This is probably due to the fact that in our formalism the pairing is not 
included explicitly which may be important in these nuclei. \\ 

\begin{figure}[hbt]
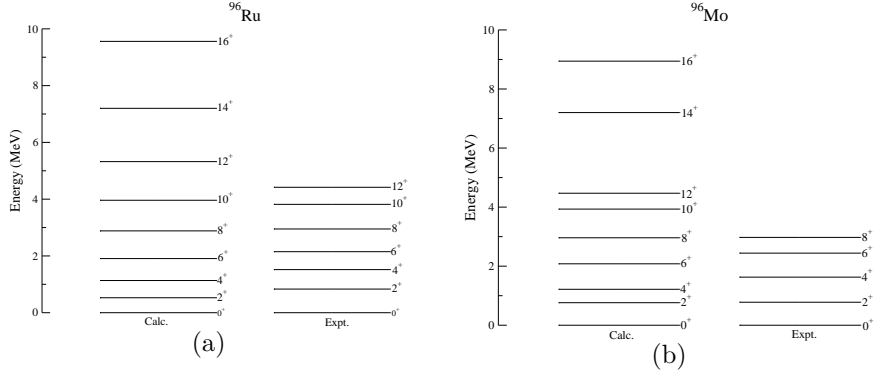

\centering
\mbox{
\subfigure[]{\includegraphics[scale=0.25, angle=-90]{plots/spectra-Ru96.epsi}}
\hspace{ 0.5cm}
\subfigure[]{\includegraphics[scale=0.26, angle=-90]{plots/spectra-Mo96.epsi}}
}
\caption{Energy spectra for $^{96}$Ru and $^{96}$Mo  nuclei
The experimental values are taken from Refs. \cite{nds96}}
\label{fig:spec_ru96}
\end{figure}

\begin{figure}[hbt]
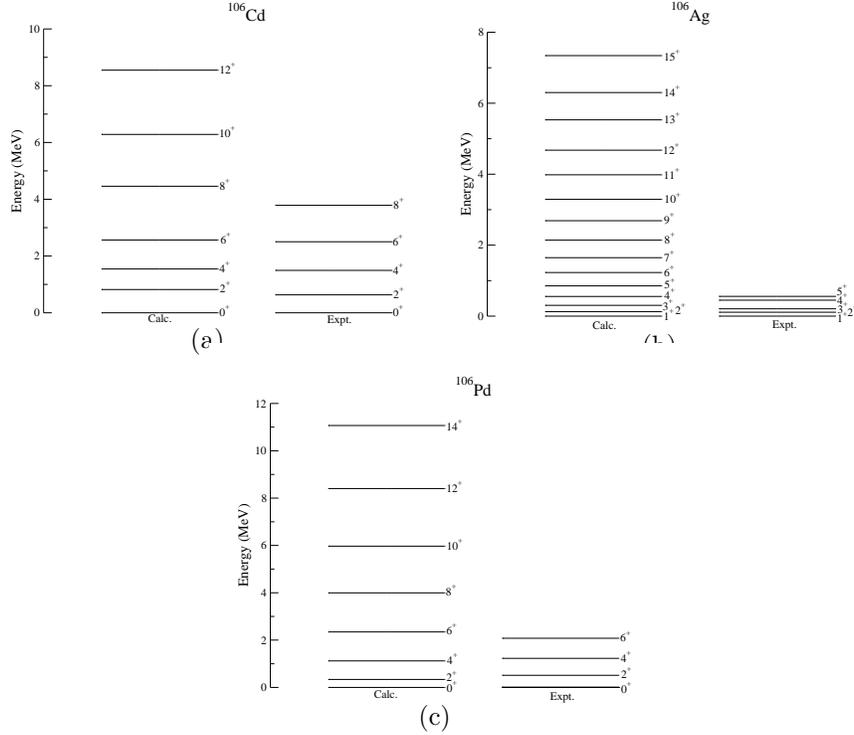

\centering
\mbox{
\subfigure[]{\includegraphics[scale=0.25, angle=-90]{plots/spectra-Cd106.epsi}}
\hspace{ 0.5cm}
\subfigure[]{\includegraphics[scale=0.25, angle=-90]{plots/spectra-Ag106.epsi}}
}
\subfigure[]{\includegraphics[scale=0.25, angle=-90]{plots/spectra-Pd106.epsi}}
\caption{Energy spectra for $^{106}$Cd, $^{106}$Ag and $^{106}$Pd 
nucleus. The experimental values are taken from Ref. \cite{nds106}}
\label{fig:spec_cd106}
\end{figure}

In Figs.~\ref{fig:spec_ru96} and \ref{fig:spec_cd106} we have shown the calculated 
ground band in comparison with 
experimentally available results for the nuclei $^{96}$Ru,$^{96}$Mo, $^{106}$Cd, $^{106}$Ag
and $^{106}$Pd. Our theoretically calculated results are in reasonably good agreement
with the experimentally known values. 

\subsection{Electromagnetic Properties}

We have calculated the reduced transition moments, quadrupole moments and magnetic dipole moments.
These values are presented in Tables~\ref{tab:tab-be2} and \ref{tab:tab-qsmu}.

In Table~\ref{tab:tab-be2} we have shown the calculated $B(E2;0^{+}\rightarrow2^{+})$ for the
ground band of the even-even parent and daughter nuclei. The values are calculated for the 
effective charges $e_{p}=1+e_{ff}$ and  $e_{n}=e_{ff}$ for protons and neutrons respectively 
where $e_{ff}$ varies for three different values 0.4, 0.5 and 0.6. Comparison between 
theoretical and experimental values are also made in this table and we obtained a reasonable 
agreement for all the  nuclei studied here except for $^{116}$Cd, $^{116}$Sn and $^{124}$Te. 
But still our calculated results are much better reproduced than the values predicted by
some other models e. g.
Woods–Saxon model and Hartree–Fock + BCS calculations with the Skyrme SIII force (see Ref.~\cite{ram01}
for the values calculated with these model) for these isotopes.
The nuclei upto $A=82$ , the values are better reproduced by taking $e_{ff}=0.5$ where as for 
the nuclei above this mass range the values those are calculated with $e_{ff}=0.6$ are more 
close to the experimental results.

The calculated quadrupole moments are given in columns 3-5 of Table~\ref{tab:tab-qsmu}. The 
values are calculated with the same effective charges as in case of B(E2). Experimental 
values are not available for $^{78}$Kr, $^{78}$Br, $^{82}$Kr, $^{96}$Tc, $^{100}$Tc, 
$^{106}$Ag, $^{130}$I and $^{130}$Xe. In the last two columns of Table~\ref{tab:tab-qsmu} 
presents the theoretically calculated and experimental results for magnetic dipole moments 
respectively. The spin and parity of the corresponding states are given in  
Table~\ref{tab:tab-qsmu}. Our calculations correctly reproduce the sign and the values are
reasonably agree with that of experimentally observed for most of the nuclei. For $^{82}$Kr 
and $^{106}$Pd the calculated values are different from experimental data by 0.6 and 0.8 respectively.

\begin{longtable}{llllllllll}
\caption{Comparison of calculated and experimentally
observed reduced transition probabilities $B(E2$:$0^{+}\rightarrow 2^{+})$.
Here $B(E2)$ are calculated for effective charge $e_{p}=1+e_{eff}$ and $%
e_{n}=e_{eff}$. In the last column the values marked by $^{*}$ are average $B(E2)$ values
 from Ref.~\cite{ram01}}\\
\hline\hline
Parent  & \multicolumn{4}{c}{$B(E2$:$0^{+}\rightarrow 2^{+})$ ($e^{2}$b$^{2}$%
)} & Daughter  & \multicolumn{4}{c}{$B(E2$:$0^{+}\rightarrow 2^{+})$ ($e^{2}$%
b$^{2}$)} \\ \cline{2-5} \cline{7-10}
Nucleus & \multicolumn{3}{c}{Theory} & Experiment{\cite{ram01}} & Nucleus & 
\multicolumn{3}{c}{Theory} & Experiment{\cite{ram01}} \\ 
& \multicolumn{3}{c}{$e_{eff}$} &  &  & \multicolumn{3}{c}{$e_{eff}$} & 
\multicolumn{1}{l}{} \\ \cline{2-4} \cline{7-9}
& 0.40 & 0.50 & 0.60 &  &  & 0.40 & 0.50 & 0.60 & \multicolumn{1}{l}{} \\ 
\hline
\endfirsthead
\multicolumn{6}{c}{{\tablename} \thetable{} -- Continued} \\
\hline
Parent  & \multicolumn{4}{c}{$B(E2$:$0^{+}\rightarrow 2^{+})$ ($e^{2}$b$^{2}$%
)} & Daughter  & \multicolumn{4}{c}{$B(E2$:$0^{+}\rightarrow 2^{+})$ ($e^{2}$%
b$^{2}$)} \\ \cline{2-5} \cline{7-10}
Nucleus & \multicolumn{3}{c}{Theory} & Experiment{\cite{ram01}} & Nucleus & 
\multicolumn{3}{c}{Theory} & Experiment{\cite{ram01}} \\ 
& \multicolumn{3}{c}{$e_{eff}$} &  &  & \multicolumn{3}{c}{$e_{eff}$} & 
\multicolumn{1}{l}{} \\ \cline{2-4} \cline{7-9}
& 0.40 & 0.50 & 0.60 &  &  & 0.40 & 0.50 & 0.60 & \multicolumn{1}{l}{} \\ 
\hline
\endhead
\hline
\multicolumn{6}{l}{{Continued on next page\ldots}} \\
\endfoot
\endlastfoot

$^{76}$Ge & 0.195 & 0.242 & 0.295 & 0.268$\pm 0.008^{*}$ & $^{76}$Se & 0.311
& 0.400 & 0.501 & \multicolumn{1}{l}{0.420$\pm 0.010^{*}$} \\ 
&  &  &  & 0.230$\pm 0.035$ &  &  &  &  & \multicolumn{1}{l}{0.39$\pm 0.26$}
\\ 
&  &  &  & 0.290$\pm 0.030$ &  &  &  &  & \multicolumn{1}{l}{0.43$\pm 0.06$}
\\ 
$^{78}$Kr & 0.521 & 0.650 & 0.793 & 0.633$\pm 0.039^{*}$ & $^{78}$Se & 0.200
& 0.254 & 0.314 & \multicolumn{1}{l}{0.335$\pm 0.009^{*}$} \\ 
&  &  &  & 0.54$\pm 0.13$ &  &  &  &  & \multicolumn{1}{l}{0.327$\pm 0.007$}
\\ 
&  &  &  & 0.686$\pm 0.030$ &  &  &  &  & \multicolumn{1}{l}{0.40$\pm 0.07$}
\\ 
$^{82}$Se & 0.161 & 0.197 & 0.237 & 0.185$\pm 0.002^{*}$ & $^{82}$Kr & 0.155
& 0.192 & 0.232 & \multicolumn{1}{l}{0.223$\pm 0.010^{*}$} \\ 
&  &  &  & 0.213$\pm 0.019$ &  &  &  &  & \multicolumn{1}{l}{0.19$\pm 0.05$}
\\ 
&  &  &  & 0.179$\pm 0.019$ &  &  &  &  & \multicolumn{1}{l}{0.225$\pm 0.009$%
} \\ 
$^{96}$Mo & 0.138 & 0.175 & 0.216 & 0.271$\pm $0.005$^{*}$ & $^{96}$Ru & 
0.143 & 0.176 & 0.212 & \multicolumn{1}{l}{0.251$\pm $0.010$^{*}$} \\ 
&  &  &  & 0.310$\pm $0.047 &  &  &  &  & \multicolumn{1}{l}{0.268$\pm $0.032
} \\ 
&  &  &  & 0.302$\pm $0.039 &  &  &  &  & \multicolumn{1}{l}{0.236$\pm $0.007
} \\ 
$^{100}$Mo & 0.306 & 0.394 & 0.493 & 0.516$\pm $0.010$^{*}$ & $^{100}$Ru & 
0.306 & 0.390 & 0.484 & \multicolumn{1}{l}{0.490$\pm $0.005$^{*}$} \\ 
&  &  &  & 0.511$\pm $0.009 &  &  &  &  & \multicolumn{1}{l}{0.493$\pm $0.003
} \\ 
&  &  &  & 0.526$\pm $0.026 &  &  &  &  & \multicolumn{1}{l}{0.494$\pm $0.006
} \\ 
$^{106}$Cd & 0.464 & 0.598 & 0.749 & 0.410$\pm $0.020$^{*}$ & $^{106}$Pd & 
0.439 & 0.552 & 0.679 & \multicolumn{1}{l}{0.660$\pm $0.035$^{*}$} \\ 
&  &  &  & 0.47$\pm $0.05 &  &  &  &  & \multicolumn{1}{l}{0.689$\pm $0.37}
\\ 
&  &  &  & 0.384$\pm $0.005 &  &  &  &  & \multicolumn{1}{l}{0.59$\pm $0.009}
\\ 
$^{110}$Pd & 0.498 & 0.617 & 0.750 & 0.870$\pm $0.040$^{*}$ & $^{110}$Cd & 
0.484 & 0.617 & 0.767 & \multicolumn{1}{l}{0.450$\pm $0.020$^{*}$} \\ 
&  &  &  & 0.78$\pm $0.12 &  &  &  &  & \multicolumn{1}{l}{0.504$\pm $0.040}
\\ 
&  &  &  & 0.820$\pm $0.080 &  &  &  &  & \multicolumn{1}{l}{0.447$\pm $0.035
} \\ 
$^{116}$Cd & 0.197 & 0.237 & 0.281 & 0.560$\pm 0.020^{*}$ & $^{116}$Sn & 
0.064 & 0.084 & 0.106 & \multicolumn{1}{l}{0.209$\pm 0.006^{*}$} \\ 
&  &  &  & 0.501$\pm 0.047$ &  &  &  &  & \multicolumn{1}{l}{0.183$\pm 0.037$%
} \\ 
&  &  &  & 0.608$\pm 0.030$ &  &  &  &  & \multicolumn{1}{l}{0.165$\pm 0.030$%
} \\ 
$^{124}$Sn & 0.0875 & 0.102 & 0.125 & 0.1160$\pm 0.0040^{*}$ & $^{124}$Te & 
0.107 & 0.132 & 0.160 & \multicolumn{1}{l}{0.568$\pm 0.006^{*}$} \\ 
&  &  &  & 0.140$\pm 0.030$ &  &  &  &  & \multicolumn{1}{l}{0.39$\pm 0.08$}
\\ 
&  &  &  & 0.188$\pm 0.013$ &  &  &  &  & \multicolumn{1}{l}{0.539$\pm 0.028$%
} \\ 
$^{130}$Te & 0.149 & 0.189 & 0.229 & 0.295$\pm $0.007$^{*}$ & $^{130}$Xe & 
0.297 & 0.377 & 0.465 & \multicolumn{1}{l}{0.65$\pm $0.05$^{*}$} \\ 
&  &  &  & 0.290$\pm $0.011 &  &  &  &  & \multicolumn{1}{l}{0.631$\pm $0.048
} \\ 
&  &  &  & 0.260$\pm $0.050 &  &  &  &  & \multicolumn{1}{l}{0.640$\pm $0.160
} \\ \hline\hline

\label{tab:tab-be2}
\end{longtable}



\noindent 
\begin{longtable}{llllllcll}
\caption{Comparison of calculated and experimentally
observed static quadrupole moments \ $Q(J^{\pi })$ and magnetic dipole
moments $\mu (J^{\pi })$ . Here $Q(J^{\pi })$ are calculated for effective
charge $e_{p}=1+e_{eff}$ and $e_{n}=e_{eff}$.}\\
\hline\hline
Nucleus & $J^{\pi }$ & \multicolumn{4}{c}{$Q(J^{\pi })$ (eb)} &  & 
\multicolumn{2}{c}{$\mu (J^{\pi })$ (nm)} \\ \cline{3-6}\cline{8-9}
&  & \multicolumn{3}{c}{Theory} & Experiment\cite{sto05} & 
\multicolumn{1}{l}{} & Theory & Experiment\cite{sto05} \\ 
&  & \multicolumn{3}{c}{$e_{eff}$} &  & \multicolumn{1}{l}{} &  &  \\ 
\cline{3-5}
&  & 0.40 & 0.50 & 0.60 &  & \multicolumn{1}{l}{} &  &  \\ \hline
\endfirsthead
\multicolumn{6}{c}{{\tablename} \thetable{} -- Continued} \\
\hline
Nucleus & $J^{\pi }$ & \multicolumn{4}{c}{$Q(J^{\pi })$ (eb)} &  & 
\multicolumn{2}{c}{$\mu (J^{\pi })$ (nm)} \\ \cline{3-6}\cline{8-9}
&  & \multicolumn{3}{c}{Theory} & Experiment\cite{sto05} & 
\multicolumn{1}{l}{} & Theory & Experiment\cite{sto05} \\ 
&  & \multicolumn{3}{c}{$e_{eff}$} &  & \multicolumn{1}{l}{} &  &  \\ 
\cline{3-5}
&  & 0.40 & 0.50 & 0.60 &  & \multicolumn{1}{l}{} &  &  \\ 
\hline
\endhead
\hline
\multicolumn{6}{l}{{Continued on next page\ldots}} \\
\endfoot
\endlastfoot

$^{76}$Ge & 2$^{+}$ & -0.166 & -0.167 & -0.168 & -0.19$\pm 0.06$ & 
\multicolumn{1}{l}{} & 0.582 & 0.84$\pm 0.05$ \\ 
&  &  &  &  &  & \multicolumn{1}{l}{} &  & 0.67$\pm 0.08$ \\ 
&  &  &  &  &  & \multicolumn{1}{l}{} &  & 0.56$\pm 0.12$ \\ 

$^{76}$As & 2$^{-}$ & 0.249 & 0.278 & 0.307 & 7$\pm 8$ & \multicolumn{1}{l}{}
& -0.775 & -0.906$\pm 0.005$ \\ 
&  &  &  &  &  & \multicolumn{1}{l}{} &  & (-)0.9028$\pm 0.0010$ \\ 

$^{76}$Se & 2$^{+}$ & -0.423 & -0.479 & -0.563 & -0.34$\pm 0.07$ & 
\multicolumn{1}{l}{} & 0.637 & 0.81$\pm 0.05$ \\ 
&  &  &  &  &  & \multicolumn{1}{l}{} &  & +0.8$\pm 0.2$ \\ 

$^{78}$Kr & 2$^{+}$ & -0.654 & -0.730 & -0.806 &  & \multicolumn{1}{l}{} & 
0.998 & +0.86$\pm 0.02$ \\ 
&  &  &  &  &  & \multicolumn{1}{l}{} &  & +1.08$\pm 0.10$ \\ 

$^{78}$Br & 1$^{+}$ & 0.217 & 0.243 & 0.268 &  & \multicolumn{1}{l}{} & 0.255
& 0.13$\pm 0.03$ \\ 

$^{78}$Se & 2$^{+}$ & -0.405 & -0.456 & -0.508 & -0.26$\pm 0.09$ & 
\multicolumn{1}{l}{} & 0.541 & 0.77$\pm 0.05$ \\ 
&  &  &  &  &  & \multicolumn{1}{l}{} &  & +0.8$\pm 0.2$ \\ 

$^{82}$Se & 2$^{+}$ & -0.365 & -0.405 & -0.444 & -0.22$\pm 0.07$ & 
\multicolumn{1}{l}{} & 0.937 & 0.99$\pm 0.06$ \\ 
&  &  &  &  &  & \multicolumn{1}{l}{} &  & 0.9$\pm 0.3$ \\ 

$^{82}$Br & 5$^{-}$ & 0.659 & 0.735 & 0.812 & 0.69$\pm 0.02$ & 
\multicolumn{1}{l}{} & 1.216 & +1.6270$\pm 0.0005$ \\ 
&  &  &  &  & 0.748$\pm 0.10$ & \multicolumn{1}{l}{} &  &  \\ 

$^{82}$Kr & 2$^{+}$ & -0.706 & -0.796 & -0.866 &  & \multicolumn{1}{l}{} & 
0.202 & 0.80$\pm 0.03$ \\ 

$^{96}$Mo & 2$^{+}$ & -0.334 & -0.376 & -0.418 & -0.20$\pm 0.08$ & 
\multicolumn{1}{l}{} & +0.91 & +0.79$\pm 0.06$ \\ 
&  &  &  &  & or +0.04$\pm 0.08$ & \multicolumn{1}{l}{} &  &  \\ 

$^{96}$Tc & 7$^{+}$ & 0.534 & 0.605 & 0.676 & - & \multicolumn{1}{l}{} & 
4.618 & 5.09$\pm 0.05$ \\ 
&  &  &  &  &  & \multicolumn{1}{l}{} &  & +5.04$\pm 0.08$ \\ 

$^{96}$Ru & 2$^{+}$ & -0.338 & -0.375 & -0.412 & -0.13$\pm 0.09$ & 
\multicolumn{1}{l}{} & 1.482 & - \\ 
&  &  &  &  & -0.1$\pm 0.2$ & \multicolumn{1}{l}{} &  &  \\ 
&  &  &  &  & -0.2$\pm 0.3$ & \multicolumn{1}{l}{} &  &  \\ 

$^{100}$Mo & 2$^{+}$ & -0.598 & -0.665 & -0.738 & -0.42$\pm $0.09 & 
\multicolumn{1}{l}{} & +1.05 & +0.94$\pm 0.07$ \\ 
&  &  &  &  & -0.39$\pm $0.08 & \multicolumn{1}{l}{} &  & +0.7$\pm 0.4$ \\ 

$^{100}$Tc & 1$^{+}$ & 0.214 & 0.237 & 0.260 & - & \multicolumn{1}{l}{} & 
-1.688 & - \\ 

$^{100}$Ru & 2$^{+}$ & -0.486 & -0.55 & -0.614 & -0.54$\pm $0.07 & 
\multicolumn{1}{l}{} & 0.98 & +1.02$\pm 0.13$ \\ 
&  &  &  &  & -0.40$\pm $0.12 & \multicolumn{1}{l}{} &  &  \\ 
&  &  &  &  & -0.43$\pm $0.07 & \multicolumn{1}{l}{} &  &  \\ 

$^{106}$Cd & 2$^{+}$ & -0.575 & -0.584 & -0.654 & -0.28$\pm 0.08$ & 
\multicolumn{1}{l}{} & 0.749 & +0.8$\pm 0.2$ \\ 

$^{106}$Ag & 1$^{+}$ & 0.189 & 0.213 & 0.238 & - & \multicolumn{1}{l}{} & 
+2.348 & +2.9$\pm 0.2$ \\ 

$^{106}$Pd & 2$^{+}$ & -0.599 & -0.672 & -0.745 & -0.56$\pm 0.08$ & 
\multicolumn{1}{l}{} & 0.711 & +0.80$\pm 0.04$ \\ 
&  &  &  &  & -0.51$\pm 0.07$ & \multicolumn{1}{l}{} &  &  \\ 

$^{110}$Pd & 2$^{+}$ & -0.536 & -0.597 & -0.658 & -0.62$\pm $0.06 & 
\multicolumn{1}{l}{} & 1.122 & -0.47$\pm 0.03$ \\ 
&  &  &  &  & -0.70$\pm $0.06 & \multicolumn{1}{l}{} &  & -0.55$\pm 0.08$ \\ 
&  &  &  &  & 0.74$\pm $0.06 & \multicolumn{1}{l}{} &  &  \\ 

$^{110}$Ag & 1$^{+}$ & 0.193 & 0.219 & 0.244 & 0.24$\pm 0.12$ & 
\multicolumn{1}{l}{} & 2.715 & 2.7271$\pm 0.0008$ \\ 

$^{110}$Cd & 2$^{+}$ & -0.515 & -0.549 & -0.654 & -0.40$\pm $0.04 & 
\multicolumn{1}{l}{} & 0.637 & +0.57$\pm 0.11$ \\ 
&  &  &  &  & -0.39$\pm $0.05 & \multicolumn{1}{l}{} &  & +0.56$\pm 0.10$ \\ 
&  &  &  &  & -0.36$\pm $0.08 & \multicolumn{1}{l}{} &  & 0.62$\pm 0.14$ \\ 

$^{116}$Cd & 2$^{+}$ & -0.334 & -0.365 & -0.396 & -0.42$\pm 0.04$ &  & +1.416
& +0.60$\pm 0.14$ \\ 
&  &  &  &  & -0.42$\pm 0.08$ &  &  &  \\ 
&  &  &  &  & -0.64$\pm 0.12$ &  &  &  \\ 

$^{116}$In & 1$^{+}$ & 0.300 & 0.335 & 0.370 & 0.11$\pm 0.01$ &  & 2.7645 & 
2.7876 \\ 

$^{116}$Sn & 2$^{+}$ & -0.203 & -0.230 & -0.258 & -0.17$\pm 0.04$ &  & 0.358
& -0.3$\pm 0.2$ \\ 

$^{124}$Sn & 2$^{+}$ & -0.246 & -0.276 & -0.305 & 0.0$\pm 0.2$ & 
\multicolumn{1}{l}{} & -0.246 & -0.3$\pm 0.2$ \\ 

$^{124}$Sb & 3$^{-}$ & 0.580 & 0.653 & 0.725 & 1.20$\pm 0.02$ & 
\multicolumn{1}{l}{} & 1.624 & +1.9$\pm 0.4$ \\ 

$^{124}$Te & 2$^{+}$ & 0.286 & 0.318 & 0.350 & -0.45$\pm 0.05$ & 
\multicolumn{1}{l}{} & 0.649 & +0.56$\pm 0.06$ \\ 
&  &  &  &  &  & \multicolumn{1}{l}{} &  & +0.66$\pm 0.06$ \\ 
&  &  &  &  &  & \multicolumn{1}{l}{} &  & +0.62$\pm 0.08$ \\ 

$^{130}$Te & 2$^{+}$ & -0.345 & -0386 & -0.428 & -0.15$\pm $0.10 & 
\multicolumn{1}{l}{} & 0.420 & +0.58$\pm 0.10$ \\ 
&  &  &  &  &  & \multicolumn{1}{l}{} &  & +0.66$\pm 0.16$ \\ 

$^{130}$I & 5$^{+}$ & 0.967 & 1.085 & 1.202 & - & \multicolumn{1}{l}{} & 
3.708 & 3.349$\pm 0.007$ \\ 

$^{130}$Xe & 2$^{+}$ & -0.491 & -0.552 & -0.614 & - & \multicolumn{1}{l}{} & 
0.611 & +0.67$\pm 0.02$ \\ 
&  &  &  &  &  & \multicolumn{1}{l}{} &  & +0.76$\pm 0.14$ \\ 
&  &  &  &  &  & \multicolumn{1}{l}{} &  & +0.62$\pm 0.08$ \\ 
\hline\hline

\label{tab:tab-qsmu}
\end{longtable}



 
\section{Conclusion}
To summarize, we have quantified the calculated deformed Hartree-Fock wave functions
by comparing the calculated values with experimental data for a number of nuclear
properties like band spectra, reduced E2 transition matrix elements, quadrupole
moments and magnetic dipole moments of nuclei involved in double beta decay processes.
A reasonable agreement between calculated and experimentally observed quantities
make us confident about the reliability of the deformed few body wave functions obtained
in our microscopic self-consistent calculations. These wave functions will further be
employed for nuclear transition matrix elements calculations of double beta decay transitions.
 
\begin{acknowledgments}
CRP acknowledges the support of the
Department of Science and Technology, India (DST Project SR/S2/HEP-37/2008)
during this work.
\end{acknowledgments}

\end{document}